\documentclass[reprint,aps,prb,superscriptaddress,notitlepage]{revtex4-2}
\usepackage{amssymb,amsmath}
\usepackage{graphicx}
\usepackage{dcolumn}
\usepackage{multirow}
\usepackage{xcolor}
\usepackage{physics}
\usepackage{hyperref}
\hypersetup{linkbordercolor={blue},colorlinks = true,linkcolor={black},citecolor={black}}
\usepackage{siunitx}
\usepackage{bm}
\usepackage{nicefrac}

\begin{document}

\title{Valley splittings in Si/SiGe quantum dots with a germanium spike in the silicon well}
\author{Thomas McJunkin}
\author{E. R. MacQuarrie}
\author{Leah Tom}
\author{S. F. Neyens}
\author{J. P. Dodson}
\author{Brandur Thorgrimsson}
\author{J. Corrigan}
\author{H. Ekmel Ercan}
\author{D. E. Savage}
\author{M. G. Lagally}
\author{Robert Joynt}
\affiliation{University of Wisconsin-Madison, Madison, WI 53706, USA}
\author{S. N. Coppersmith}
\affiliation{University of Wisconsin-Madison, Madison, WI 53706, USA}
\affiliation{University of New South Wales, Sydney, Australia}
\author{Mark Friesen}
\author{M. A. Eriksson}
\affiliation{University of Wisconsin-Madison, Madison, WI 53706, USA}

\begin{abstract}
Silicon-germanium heterostructures have successfully hosted quantum dot qubits, but the intrinsic near-degeneracy of the two lowest valley states poses an obstacle to high fidelity quantum computing. 
We present a modification to the Si/SiGe heterostructure by the inclusion of a spike in germanium concentration within the quantum well in order to increase the valley splitting. 
The heterostructure is grown by chemical vapor deposition and magnetospectroscopy is performed on gate-defined quantum dots to measure the excited state spectrum.
We demonstrate a large and widely tunable valley splitting as a function of applied vertical electric field and lateral dot confinement.
We further investigate the role of the germanium spike by means of tight-binding simulations in single-electron dots and show a robust doubling of the valley splitting when the spike is present, as compared to a standard (spike-free) heterostructure. This doubling effect is nearly independent of the electric field, germanium content of the spike, and spike location. This experimental evidence of a stable, tunable quantum dot, despite a drastic change to the heterostructure, provides a foundation for future heterostructure modifications.
\end{abstract}

\maketitle

\section{Introduction}

Low-lying valleys in the silicon conduction band are a focal point in the pursuit of gate-defined quantum dot qubits. For silicon quantum wells strained between layers of relaxed silicon-germanium, there are only two low-lying valley states, separated in energy by the the valley splitting~\cite{Ando:1982p437,Schaffler:1997p1515,Friesen:2007p115318}. While defect-free Si/SiGe heterostructures are predicted to have $> 1$~\si{\milli \electronvolt} valley splitting, interfacial disorder, growth on a substrate that is mis-cut with respect to the crystal axes, and other imperfections lead to typical valley splittings in the range of tens to hundreds of~\si{\micro \electronvolt}~\cite{Kharche:2007p092109}. For single-electron spin qubits, the valley splitting should be as large as possible~\cite{Veldhorst:2015p410,Eng:2015p1500214}. For other qubit implementations, such as the quantum-dot-hybrid qubit, the valley splitting is integral to the logical state energies, and an intermediate valley splitting of order 30~\si{\micro \electronvolt} is desirable~\cite{Maune:2012p344,Kim:2014p70,Mi:2018p161404,corrigan2020coherent}.

The valley splitting in SiGe heterostructures can be tuned by controlling the vertical electric field or lateral dot position~\cite{Boykin:2004p115,Shi:2011p233108,Hosseinkhani:2020p043180,Jones:2019p014026,Hollmann:2020p034068, Dodson:2021preprint}. 
In the first case, the applied electric field affects the wave function overlap with the SiGe barrier and therefore affects the valley splitting; however, the electric field is often bounded by material limits or control limits of the dot for qubit use. 
In the second case, the lateral position of the dot determines the disorder profiles it experiences at the quantum well interface; however, this tuning method is inherently unpredictable as the disorder itself is poorly controlled or characterized. 
Previous work modifying the Si/SiGe heterostructure by inclusion of extra germanium at the quantum well interface has highlighted the critical role of disorder in these heterostructures~\cite{Neyens:2018p243107}.

Here, we explore a new method for modifying the valley splitting by including an ultra-thin layer of silicon-germanium within the silicon quantum well. This spike in germanium concentration is placed near the upper interface of the quantum well where the two-dimensional electron gas (2DEG) is formed~\cite{Kiselev:2010pB26}. It effectively splits the quantum well into two regions, significantly affecting the shapes of the two valley wave functions and increasing the energy splitting between them. We describe the growth of this heterostructure and characterize it with scanning transmission electron microscopy (STEM). Hall bars and quantum dot devices are fabricated on the heterostructure, and magnetospectroscopy is used to measure the excited state energy spectrum of a few-electron quantum dot.
These results reveal a large and tunable valley splitting. Furthermore, they demonstrate that stable and clean quantum device behavior is achievable in the presence of large modulations in Ge concentration, even when it is positioned directly at the peak of the electron wave function in the quantum well.
In order to understand how the SiGe layer within the quantum well contributes to the valley splitting, we perform one- and two-dimensional tight-binding simulations of a single-electron quantum dot in the presence of common interfacial disorder.
We present numerical calculations showing that the presence of the ultra-thin SiGe layer in this sample increases the valley splitting by a factor of two compared to a heterostructure without this layer.
This doubling is robust against large changes in the Ge concentration of the spike, its position in the quantum well, and the vertical electric field.

This paper is organized as follows. In Sec.~\ref{sec:Exp_methods} we describe the experimental growth and measurement methods. In Sec.~\ref{sec:Exp_results} we present magnetospectroscopy results at the 1-2, 2-3, and 3-4 electron charging transitions and tuning of the excited state spectrum at the 3-4 electron charging transition. In Sec.~\ref{sec:Theory}, we present both effective mass theory and tight-binding theory of the Ge spike's effect on the valley splitting. Section~\ref{sec:conclusions} is a summary, and the Appendices present details of the magnetospectroscopy measurements and tight binding calculations.

\begin{figure*}[t]
\includegraphics[width=1.0\textwidth]{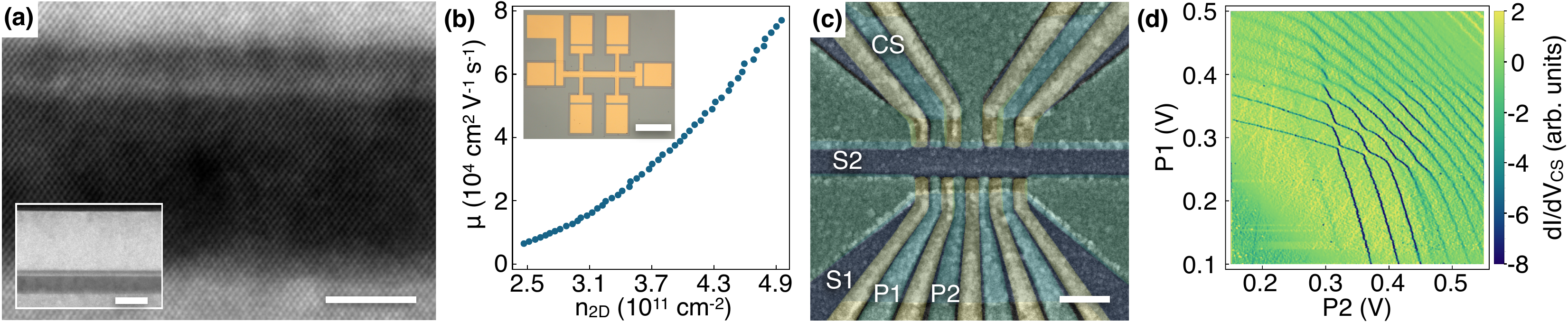}
\caption{
Heterostructure and devices. (a) High-angle annular dark-field (HAADF) images of the SiGe heterostructure characterized, acquired with a scanning transmission electron microscope (STEM). Higher brightness in the images corresponds to higher germanium concentration. The main image focuses on the silicon quantum well, with a $5$~\si{\nano \meter} scale bar. The inset shows the heterostructure from the surface to the well, with a $20$~\si{\nano \meter} scale bar. (b) Carrier density and transport mobility of a Hall bar fabricated on the heterostructure shown in (a), measured in a dilution refrigerator at $< 50$~\si{\milli \kelvin}. The inset shows an optical image of the Hall bar, with a $200$~\si{\micro \meter} scale bar. Dimensions of the Hall bar are $200$~\si{\micro \meter} long by $20$~\si{\micro \meter} wide. (c) False-colored scanning electron micrograph of the quantum dot device measured, with a $200$~\si{\nano \meter} scale bar. The different colors (blue, green, yellow) indicate different gate layers. (d) Stability diagram of a double quantum dot formed using the left two quantum dot plunger gates (P1 and P2) and charge sensing quantum dot (CS), as labeled in (c). 
}
\label{fig_device}
\end{figure*}

\section{Experimental Methods}
\label{sec:Exp_methods}

% Heterostructure measured is 18032101
The heterostructure is grown by ultra-high vacuum chemical vapor deposition (UHV-CVD) on a linearly graded SiGe alloy with a final 2~\si{\micro \meter} layer of Si$_{0.705}$Ge$_{0.295}$. Prior to heterostructure growth, the SiGe substrate is cleaned and prepared as described in Ref.~\onlinecite{Neyens:2018p243107}. The heterostructure is grown at 600~\si{\celsius} with a mixture of silane and germane gases. A $550$~\si{\nano \meter}, 29.5\si{\percent} Ge alloy layer is grown before the silicon quantum well. The main silicon well is grown, then germane is re-introduced to grow the ultra-thin layer of SiGe. The germane flow is stopped to grow the silicon layer in the quantum well above the SiGe layer, followed by a $33$~\si{\nano \meter} layer of $29.5$\si{\percent} Ge alloy. The growth is capped with a $1$~\si{\nano \meter} layer of silicon. The resulting heterostructure is shown in Fig.~\ref{fig_device}(a) in a high-angle annular dark-field (HAADF) image acquired with a scanning transmission electron microscope (STEM). The lighter colored SiGe layer within the darker colored Si quantum well is clearly visible. This spike in Ge content is approximately $1$~\si{\nano \meter} thick, creating a $\sim 1.5$~\si{\nano \meter} secondary quantum well above the $\sim 10$~\si{\nano \meter} main quantum well. In the inset, a broader view of the heterostructure shows that the location and thickness of the germanium spike is relatively constant across a wide area. 

Hall bar and quantum dot devices were fabricated simultaneously on this heterostructure. A $20$~\si{\nano \meter} layer of aluminum oxide grown by atomic layer deposition at 200~\si{\celsius} isolates the various metallic gates from the surface of the heterostructure, followed by a 15~minute, 450~\si{\celsius} forming gas anneal. All measurements were performed in a dilution refrigerator with a base temperature below $50$~\si{\milli \kelvin}. 
Measurements of the Hall bar device shown in Fig.~\ref{fig_device}(b) reveal transport mobilities in the range of 1-8$ \cross 10^4$~\si{\centi \meter}$^{2}$\si{\volt}$^{-1}$\si{\second}$^{-1}$ across an electron density range of 2.5-5$ \cross 10^{11}$~\si{\centi \meter}$^{-2}$. 
Fabrication of the quantum dots follows the procedures of Ref.~\onlinecite{Dodson:2020p505001}, using a three-layer, overlapping aluminum gate design with each layer isolated by the self-oxidation of the aluminum. This oxidation is enhanced by a 15~minute down-stream oxygen plasma ash. A false-colored scanning electron micrograph of a quadruple quantum dot device nominally identical to the one measured is shown in Fig.~\ref{fig_device}(c). The bottom half of the device shows four linearly arranged quantum dots, opposing two charge sensing quantum dots on the top half. For the work reported here, the left side of the device is used. Figure~\ref{fig_device}(d) shows a double quantum dot stability diagram in the zero-to-few electron regime of the left two dot plunger gates, P1 and P2, sensed by the charge sensor dot CS. For additional double quantum dot measurements of this device, see Ref.~\onlinecite{Zwolak:2020p034075}.

\begin{figure*}
\includegraphics[width=1.0\textwidth]{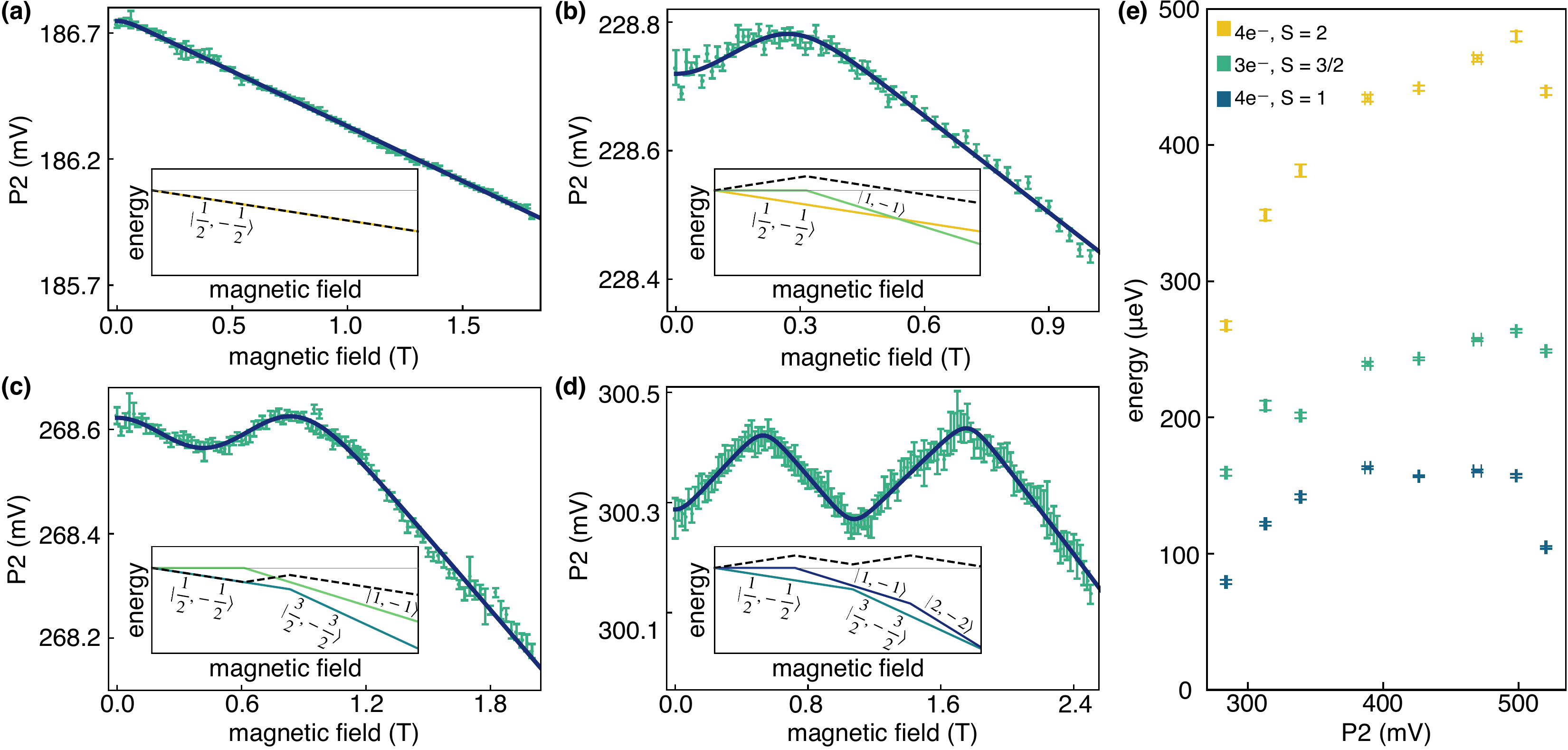}
\caption{
Magnetospectroscopy of a quantum dot. (a-d) Dependence on magnetic field of the gate voltage at which an additional electron enters the dot at the 0-1, 1-2, 2-3, and 3-4 electron charging transitions, respectively. Green points show the location of the maximum of the differentiated charge sensor signal of the electron charging transition. Dark blue lines show the computed fits to the data, as described in Appendix~\ref{A_magspec_fits}. Zero-magnetic-field spin splittings extracted from the fits in (b-d) are presented in Table~\ref{tab_magspec}.
Insets to (a-d) schematically show the changing ground spin states and resulting magnetospectroscopy curve. 
The colored lines show the energies of the ground spin states $\ket{S, m_s}$ of both electron occupations in the main figure as a function of magnetic field. 
As the spin state energies of each electron occupation differs, the energy of the electron charging transition changes with magnetic field. Subtracting the lower occupation spin energy from the higher occupation spin energy, we obtain the magnetospectroscopy curve, shown as the dashed black line. Thus, the magnetospectroscopy slope measured at any electron charging transition $n \rightarrow n+1$ is proportional to $m_s(n+1) - m_s(n)$.
(e) Magnetospectroscopy results at the 3-4 electron charging transition as a function of P2 and S1/S2 voltage tunings, as described in Sec.~\ref{sec:Exp_results}. The zero-magnetic-field excited state energies of the polarized spin states for 3 and 4 electrons are shown, labeled by their spin number, $S$. 
}
\label{fig_magspec}
\end{figure*}

In Sec.~\ref{sec:Exp_results} below we report the results of magnetospectroscopy performed on a single quantum dot, in order to characterize the valley splitting, by determining the value of the in-plane magnetic field that causes a change in the ground state spin configuration~\cite{Hada:2003p155322,Hanson:2007p1217,Lim:2009p242102,Shi:2011p233108}.
To perform magnetospectroscopy, as a function of the in-plane magnetic field, we measure the voltage on gate P2 at which electron charging transitions occur. See Appendix~\ref{A_registration} for example raw data and Appendix~\ref{A_magspec_fits} for the fitting procedures used in Fig.~\ref{fig_magspec}. We measure the $g$-factor of electrons in this heterostructure to be $g = 1.98 \pm 0.03$ (see Appendix~\ref{A_gfactor}), matching the expected value for silicon~\cite{Wilamowski:2003}.

\section{Experimental Results}
\label{sec:Exp_results}

In this section, we first discuss how the relevant energy splittings are extracted from the magnetospectroscopy measurements. Next, we report the energy splittings for different electron occupations in the dot. We then show how the splittings depend on the vertical electric field and lateral confinement in the dot by tuning the gate voltages to maintain a constant dot occupancy.

Figure~\ref{fig_magspec}(a) shows the 0-1 charge transition as a function of the in-plane magnetic field $B$, which decreases the ground state energy by $m_s g \mu_B B$, where $m_s$ is the spin projection along the magnetic field, $g$ is the electron $g$-factor, and $\mu_B$ is the Bohr magneton. The slope of the transition line is determined by the combination of the Zeeman effect and the lever arm $\alpha$ connecting the voltage on gate P2 to the chemical potential of the quantum dot. As shown in the inset, and the corresponding fit to the data points, this data is consistent with single-electron occupation of the quantum dot.

The singlet-triplet (ST) splitting can be extracted by performing the same measurement at the 1-2 charge transition, as shown in Fig.~\ref{fig_magspec}(b). For small $B$, the transition line slopes upward, because a spin of opposite sign is added to form a singlet ground state. A kink occurs in the curve when the $T_-$ state is Zeeman shifted by an amount equal to the zero-magnetic-field singlet-triplet splitting and becomes the ground state. Tracking the line labeled $\ket{1, -1}$ in the inset to zero magnetic field enables extraction of this splitting, and a value of 31.4~\si{\micro \electronvolt} is reported in Table~\ref{tab_magspec}. This is a lower bound on the single-electron valley splitting, because electron-electron interactions suppress the two-electron ST splitting from this value~\cite{Dodson:2021preprint}.

\begin{table}[t]
\setlength{\tabcolsep}{4.5pt}
\begin{tabular}{ c c c c c c }
\hline
\hline
electron&$2 e^-$&$3 e^-$&$4 e^-$&$4 e^-$&P2\\

charging&$S$=$1$&$S$=$\nicefrac{3}{2}$&$S$=$1$&$S$=$2$&$\alpha$\\

transition&(\si{\micro \electronvolt}) & (\si{\micro \electronvolt}) & (\si{\micro \electronvolt}) & (\si{\micro \electronvolt}) & (\si{\electronvolt}/\si{\volt}) \\

\hline
1$\rightarrow$2&$31.4(5)$ & - & - & - & $0.115(2)$ \\
2$\rightarrow$3&$47.9(7)$& $88.8(6)$ & - & - & $0.126(2)$ \\
3$\rightarrow$4&- & $121(1)$ & $64(1)$ & $263(4)$ & $0.17(1)$ \\
\hline
\hline
\end{tabular}
\caption
{Zero-magnetic-field excited spin state energies extracted from the magnetospectroscopy curves in Fig.~\ref{fig_magspec}(b-d). Each row corresponds to a different electron charging transition, as indicated in the first column. The second column reports the two-electron singlet-triplet splitting. The final column shows the lever arm, $\alpha$, as calculated from the magnetospectroscopy slope.
}
\label{tab_magspec}
\end{table}

Magnetospectroscopy of the 2-3 electron charging transition (Fig.~\ref{fig_magspec}(c)) displays two kinks: the first reveals the 2-electron ST splitting, but in this case at gate voltages corresponding to the 2-3 transition instead of the 1-2 transition. That is, the electric field pulling the electrons onto the ultra-thin SiGe layer and against the interface is larger, and as a result the extracted 2-electron ST splitting of 47.9~\si{\micro \electronvolt} is larger than at the 1-2 transition. The second kink reveals the energy of the zero-magnetic-field excited state $\ket{3/2, -3/2}$, as reported in Table~\ref{tab_magspec}.

Figure~\ref{fig_magspec}(d) reports magnetospectroscopy for the 3-4 electron charging transition, revealing a set of three kinks corresponding to the zero-magnetic-field energies of three excited states, as reported in Table~\ref{tab_magspec}. These excited state energies again place a lower bound on the valley splitting, in this case at yet larger electric field, corresponding to the 3-4 charge transition. 
With the larger number of electrons involved, and as shown in Appendix~\ref{A_magspec_fits}, Eqs.~(\ref{eq:v1}-\ref{eq:v2}), we can extract the valley splitting of both the ground and the first excited orbital state, in the approximation that electron-electron interactions are negligible. We find a ground state valley splitting of $57 \pm 2$~\si{\micro \electronvolt} and a valley splitting of the first orbital excited state of $78 \pm 4$~\si{\micro \electronvolt}. The first excited orbital state energy is also extracted in this approximation, which we find is $132 \pm 2$~\si{\micro \electronvolt}.

With these measurements at various electron charging transitions, we find that the valley splitting increases with electric field, consistent with results on conventional Si/SiGe heterostructures~\cite{Jones:2019p014026,Hosseinkhani:2020p043180}.
Furthermore, the 3-4 electron charging transition allows us to characterize excitations primarily associated with valleys and orbitals. 

To better understand the range of device tunings that are possible with electric fields, we perform additional magnetospectroscopy measurements at the 3-4 transition as a function of vertical electric field. Here, we simultaneously adjust the voltages on several gates, as described below, to ensure that only the electric field changes (not the dot location), while keeping the charge occupation fixed.
The quantum dot is beneath gate P2, which we make more positive in order to increase the electric field, thus pulling the dot tighter against the interface and the ultra-thin SiGe layer. Simultaneously, we make the voltages on gates S1 and S2 more negative to increase the confinement energy and adjust the barrier gate voltages to maintain reasonable tunnel rates. Appendix~\ref{A_E_tuning} shows the resulting magnetospectroscopy curves and reports COMSOL simulations which confirm that this approach increases the electric field while keeping the dot centered in the channel.

Figure~\ref{fig_magspec}(e) shows the three zero-magnetic-field excitation energies of the 3-4 electron charging transition as a function of the voltage on gate P2. We note that larger P2 gate voltage corresponds to stronger vertical electric field and stronger lateral confinement. Remarkably, each of these excitation energies can be tuned by at least $60$\si{\percent} through this tuning approach. 
Using Eq.~(\ref{eq:v1}), we estimate a tuning range of $60-144$~\si{\micro \electronvolt} for the valley splitting, as we simultaneously adjust the electric field and confinement potential. 

Previous results have shown valley splitting tunability of 15\% with a maximum value of 213~\si{\micro \electronvolt}~\cite{Hollmann:2020p034068} as well as 140\% tunability with a maximum value of 87~\si{\micro \electronvolt}~\cite{Dodson:2021preprint}. Here, we show 140\% tunability with a lower bound on the maximum valley splitting of 144~\si{\micro \electronvolt}.
Hence, we conclude that the presence of the ultra-thin SiGe layer inside the quantum well facilitates, or is at least compatible with, a large and widely tunable valley splitting in few-electron dots.

\section{Theory}
\label{sec:Theory}

\begin{figure*}[ht]
\includegraphics[width=1.0\textwidth]{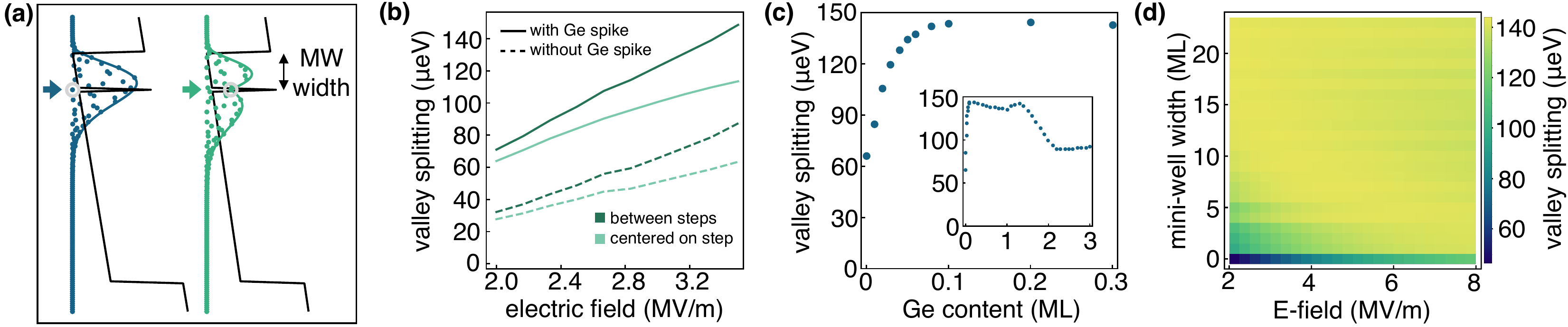}
\caption{
Results of tight-binding calculations of the valley splitting in the presence of a Ge spike, assuming realistic interfacial disorder with uniform atomic steps spaced 45~\si{\nano \meter} apart. (a) One-dimensional tight-binding calculations of wave function densities ($|\psi |^2$) for the ground state (blue circles) and the first excited state (green circles) of an electron in a 12.9~nm quantum well and vertical electric field of $5.5$~\si{\mega \volt / \meter}, with a single-atom barrier positioned 19 monolayers below the top of the quantum well. Solid lines show the density envelopes, computed as described in Appendix~\ref{A_TB_functions}, while the data points are obtained using a tight-binding theory, as described in Sec.~\ref{sec:TB_theory}. 
(b) The valley splitting of the lowest orbital of a quantum dot, where the quantum dot is placed halfway between two steps (darker lines) or centered at a step (lighter lines). The solid lines are calculated for a heterostructure modeled after the experimental growth results shown in Fig.~\ref{fig_device}(a), while the dashed lines are calculated for a standard heterostructure with no added germanium in the silicon well. The addition of a germanium spike roughly doubles the valley splitting. 
(c) Tight-binding calculations of the valley splitting in a quantum dot with a $3$~\si{\mega \volt / \meter} vertical electric field, centered halfway between two steps in a heterostructure with a monolayer (ML) spike in germanium located 20 ML below the upper interface of the quantum well. The main figure shows the valley splitting as a function of germanium content. The inset shows the same for higher germanium content, with contents greater than 1 achieved by increasing the spike thickness through additional monolayers closer to the upper interface of the well. (d) Valley splitting as a function of the location of the Ge spike and the applied electric field for a single ML spike of $30$\si{\percent} germanium. The location of the Ge spike is defined in terms of the width of the `mini-well' created between the Ge spike and the upper interface of the quantum well (shown by the black arrow in (a), labeled with MW width). For the case of zero mini-well width, the Ge spike contacts the Si$_{0.7}$Ge$_{0.3}$ capping layer, consistent with a heterostructure with no added SiGe layer. 
}
\label{fig_sim}
\end{figure*}

In order to better understand the underlying physics of the Ge spike inside the quantum well and its contribution to the valley splitting, we perform effective mass and tight-binding calculations with and without this additional SiGe layer. These calculations are performed in single-electron quantum dots.

\subsection{Effective Mass Theory}

We first begin with an effective mass theory interpretation of the valley splitting due to a single atom barrier within the quantum well. Valley splitting is determined by the interactions between electronic wave functions and barrier interfaces in the quantum well. For weak valley-orbit interactions, the two low-energy eigenfunctions $\psi$ can be approximately written as $\sqrt{2}F({\bf r})\cos(k_0z+\varphi)$ and $\sqrt{2}F({\bf r})\sin(k_0z+\varphi)$, where $F({\bf r})$ is the effective mass envelope function, $k_0$ is the position of the conduction-band valley minimum in reciprocal space, $z$ is the position coordinate in the $[001]$ growth direction, and $\varphi$ is a fixed phase that maximizes the valley splitting~\cite{Friesen:2007p115318}. 
For a single-atom barrier potential at $z = z_s$, the variable $\varphi$ is determined such that the ground-state wave function has a node at the barrier, $\cos(k_0 z_s + \varphi) = 0$, while the excited state wave function takes its maximum value, such that $\sin(k_0z_s+\varphi)=1$. 
The resulting potential energy difference for these two eigenstates is large, and can be further enhanced by positioning the barrier at the peak of the envelope function, yielding a very large valley splitting. 
For a more complete description of how the wave function envelopes differ, the valley-orbit coupling induced by the single-atom barrier should be taken into account, as described in Appendix~\ref{A_TB_functions}. The envelopes of the two valley states are plotted in Fig.~\ref{fig_sim}(a) and differ significantly, indicating an extraordinary level of mixing between the orbital and valley degrees of freedom (valley-orbit coupling). Large valley-orbit coupling occurs because the ground-state wave function has a node with $\psi=0$ at the position of the barrier (blue arrow and gray circle), while the excited-state wave function is maximized at the same location (green arrow and gray circle).

\subsection{Tight-Binding Theory}
\label{sec:TB_theory}

We investigate the effect of the germanium spike on the excited state spectrum using a tight-binding model to calculate single-electron valley splittings~\cite{Boykin:2004p115,Boykin:2004p165325}.
We first consider the one-dimensional Ge spike geometry shown in Fig.~\ref{fig_sim}(a). The resulting wave function densities for the ground and excited valley states are plotted as discrete points in Fig.~\ref{fig_sim}(a), showing good agreement with the envelopes calculated in Appendix~\ref{A_TB_functions}.

We then perform two-dimensional tight-binding simulations using the method described in Ref.~\cite{Abadillo-Uriel:2018p165438} to investigate how realistic disorder in the heterostructure affects the single-electron valley splitting.
The electric fields applied in the experiments are estimated using COMSOL, as described in Appendix~\ref{A_E_tuning}. We introduce atomistic disorder at the quantum well interfaces by assuming a uniform step density of 45~\si{\nano \meter}/step, which allows us to approximately match the valley splitting results in the simulations to the lower bounds of the valley splittings calculated from Fig.~\ref{fig_magspec}(e). Similarly, the lower bound of the orbital splitting, calculated from Fig.~\ref{fig_magspec}(e), is adopted as the lateral parabolic confinement $\hbar \omega$ in the tight-binding model. As noted above, this correspondence is not accurate in the presence of strong electron-electron interactions; however, it allows us to establish an approximate working regime for our simulations. 

The simulations are performed for both a standard SiGe heterostructure and a heterostructure with an added realistic spike in germanium, whose shape profile is estimated from the STEM results in Fig.~\ref{fig_device}(a). These results are presented in Fig.~\ref{fig_sim}(b). Because the location of the dot with respect to the background of atomic steps influences the valley-orbit coupling, we perform two sets of simulations: the first with the dot centered halfway between two steps, and the second with the dot centered directly on a step. For the lowest orbital, the addition of the germanium spike within the quantum well results in a doubling of the valley splitting compared to a standard heterostructure, as shown in Fig.~\ref{fig_sim}(b). This doubling is approximately independent of both the quantum dot position and the applied electric field, indicating a rather robust effect.
Additionally, the simulated increasing lateral confinement with increasing electric field results in an increasing valley splitting for all simulated interfaces.

To investigate further how a spike in germanium within the quantum well affects the valley splitting, we also vary the germanium concentration, spike location, and electric field in tight-binding simulations, for the case of a mono-atomic-layer spike. Figure~\ref{fig_sim}(c) shows the valley splitting dependence on germanium content within the spike. For a mono-atomic-layer spike, we observe that the doubling of the valley splitting is already achieved with less than $10$\si{\percent} germanium. This remains true up to a thickness of 1.5 monolayers of pure germanium, as shown in the inset. Figure~\ref{fig_sim}(d) shows the valley splitting dependence on both the vertical location of a $30$\si{\percent} Ge spike and the applied electric field. Here, we define the spike position in terms of the `mini-well' width, referring to the number of atomic layers in the portion of the quantum well above the top of the spike and shown by the black arrow in Fig.~\ref{fig_sim}(a). 
We note that a mini-well of width zero corresponds to a typical heterostructure without the Ge spike.
Remarkably, the observed doubling in the valley splitting is fairly resilient to both spike positioning and vertical electric field, over its entire range. For the case of zero mini-well width, the valley splitting shows a strong dependence on vertical electric field as more of the wave function is pulled into the alloy. Once the Ge spike is separated by several monolayers from the interface, however, the valley splitting remains essentially constant as a function of spike position and electric field. These results suggest an enhanced valley splitting can be obtained without needing to carefully position the germanium spike. 

\section{Conclusions}
\label{sec:conclusions}

We have studied a new heterostructure containing a spike in germanium concentration in the quantum well at the approximate location of the 2DEG. Hall bars and quantum dots were successfully fabricated and measured on this structure. Magnetospectroscopy measurements were used to probe the few-electron energy splittings and explore how these splittings change with gate voltage tunings. We showed large and widely tunable few-electron energy splittings, arising from a large and tunable single-electron valley splitting.
Tight-binding simulations in a single-electron quantum dot showed that the valley splitting doubles in the presence of the germanium spike, and that this effect should be robust against typical growth imperfections. These results serve as an example that large changes to the standard Si/SiGe heterostructure still allow for stable quantum dot formation while modifying its underlying properties.

\section{Acknowledgments}

This research was sponsored in part by the Army Research Office (ARO), through Grant Number W911NF-17-1-0274 and the Vannevar Bush Faculty Fellowship program sponsored by the Basic Research Office of the Assistant Secretary of Defense for Research and Engineering and funded by the Office of Naval Research through Grant No. N00014-15-1-0029. Development and maintenance of the growth facilities used for fabricating samples was supported by DOE (DE-FG02-03ER46028). We acknowledge the use of facilities supported by NSF through the UW-Madison MRSEC (DMR-1720415) and the MRI program (DMR–1625348). The views and conclusions contained in this document are those of the authors and should not be interpreted as representing the official policies, either expressed or implied, of the Army Research Office (ARO), or the U.S. Government. The U.S. Government is authorized to reproduce and distribute reprints for Government purposes notwithstanding any copyright notation herein.

\appendix

\section{Magnetospectroscopy Data Acquisition and Registration}
\label{A_registration}

\begin{figure}
\includegraphics[width = 0.45\textwidth]{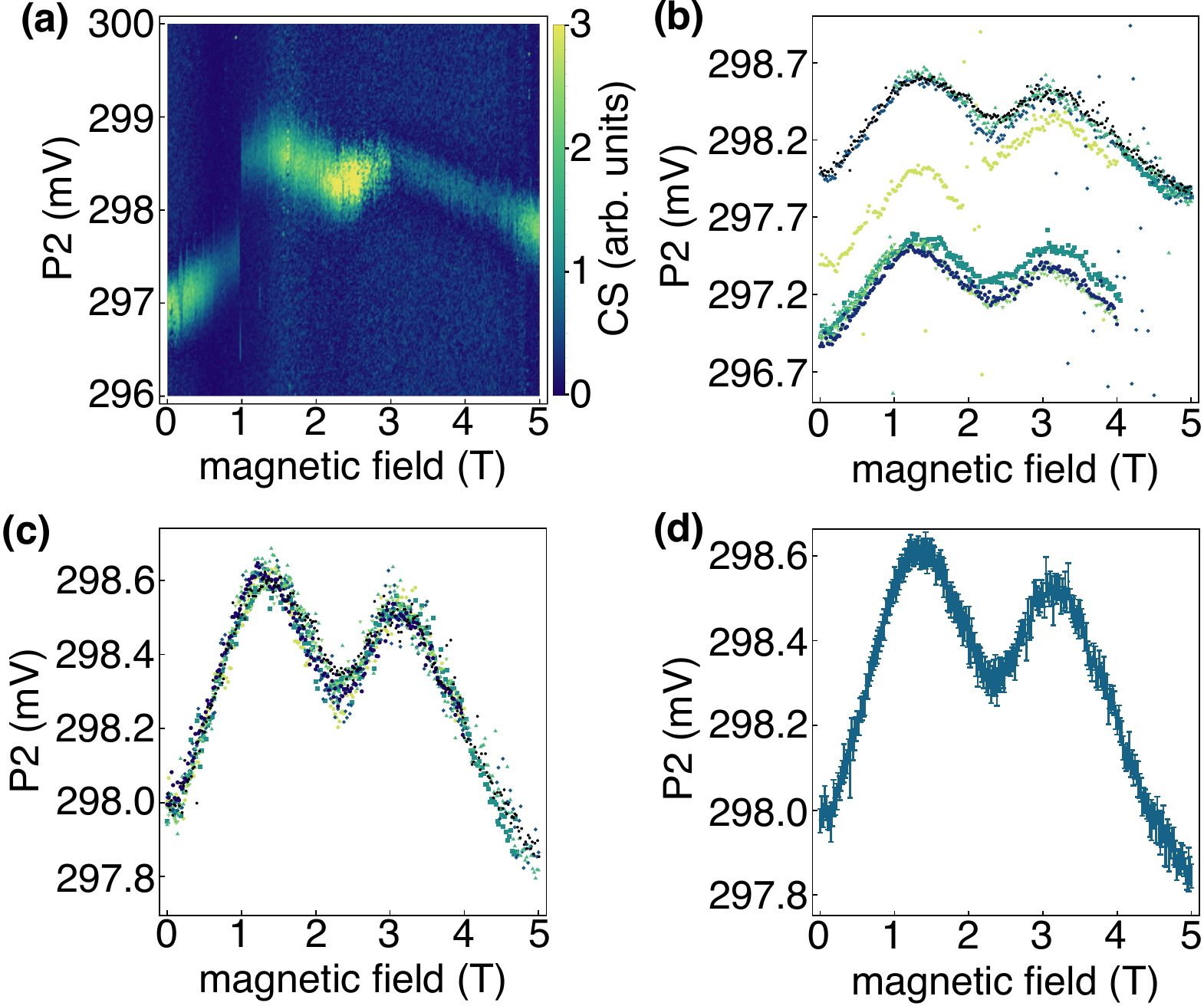}
\caption{
Registration of magnetospectroscopy data. (a) Example magnetospectroscopy dataset with characteristic `jump' near 1 Tesla. (b) Extracted peak locations of unmodified datasets. Black points are the reference scan, colored points are all other datasets at this tuning. (c) Same data as in (b), with registration shifts. (d) Final dataset, with plotted voltage error.
}
\label{figSreg}
\end{figure}

The electron charging transition of the dot formed under P2 is sensed by the charge sensor dot CS (labeled in Fig.~\ref{fig_device}(c)). A small ac voltage is applied to P2 and the current through the charge sensor is measured by a lock-in amplifier, such that electron charging transitions appear as peaks in the lock-in response, as shown in Fig.~\ref{figSreg}(a). To extract the peak location of the electron charging transition, we fit to each fixed-magnetic-field voltage scan. This fit, based on conductance through a quantum dot \cite{Kouwenhoven:1997p1384}, is
\begin{equation}
\dfrac{d I_{sensor}}{dV} = a + b*(\cosh[\frac{V-V_{peak}}{c}])^{-2},
\label{eq_CSfit}
\end{equation}
where $V_{peak}$ is the extracted peak location and $a$, $b$, and $c$ are additional fit parameters not used in this analysis. 

Due to the length of time required for a magnetospectroscopy scan ($\sim$hours) and the relative instability of the nearby charge landscape, finite shifts in the electrostatic potential of the quantum dot result in few mV shifts in the dot transition with respect to the P2 voltage during the measurement, as shown near 1~\si{\tesla} in Fig.~\ref{figSreg}(a). These shifts occur infrequently ($\sim$once per hour), allowing for repeated magnetic field sweeps to fully measure the magnetospectroscopy curve. After measurement, registration of the various scans is performed to achieve a single magnetospectroscopy curve.

To perform this registration, multiple magnetospectroscopy scans are repeatedly collected. The scan with the largest segment of data uninterrupted by a charge jump is determined and used as the registration reference. Each additional scan is registered to this reference in the magnetic field range in which they overlap (Fig.~\ref{figSreg}(b)). To calculate the voltage shift required to register each scan or a portion of a scan, the sum of the squared residuals between them is calculated, excluding the largest 10\% of residuals. This sum is then minimized by allowing for an overall voltage shift to the scan, resulting in overlapping data sets like that shown in Fig.~\ref{figSreg}(c). This process is repeated for all additional scans. A final data set is created (Fig.\ \ref{figSreg}(d)) by calculating the weighted mean at each magnetic field value, using the uncertainty in the fit for the peak location from the original scans as the weights. 

\section{Magnetospectroscopy Fitting}
\label{A_magspec_fits}

In order to fit magnetospectroscopy data, we implement a model utilizing a grand canonical ensemble, in which both the energy and the number of particles can be exchanged between a quantum dot and a reservoir. The energy states $E_i$ of the quantum dot with $N_i$ electrons can be written as a function of the charging energy, $E_c$, the Zeeman energy of the spins, $E_Z = m_s g \mu_B B$, the zero-magnetic-field energy of the excited spin states, $E_{ex}$, and the electrostatic energy imparted by the gate voltage, $V$, using the lever-arm conversion factor $\alpha$~\cite{vanderWiel:2003p1}:
\begin{equation}
E_i = \frac{1}{2}{N_i}^2 E_c + E_{Z} + E_{ex} - N_i \alpha V.
\label{Eq_Ei}
\end{equation}
To calculate the average number of particles, we use
\begin{equation}
\langle N \rangle = \frac{\sum_{i}^{} N_i g_i \exp{(\beta (N_i \mu - E_i))}}{\sum_{i}^{} g_i \exp{(\beta (N_i \mu - E_i))}},
\label{Eq_avgN}
\end{equation}
where $g_i$ is the multiplicity of the $i^{th}$ state, $\beta$ is the Boltzmann factor $1/k_b T$ and $\mu$ is the chemical potential of the dot. To analyze the charging transition from $N$ to $N+1$ electrons, we note that the peak in conduction (or peak in differentiated charge sensor signal) corresponds to an average electron occupation of $N+1/2$. Therefore, by equating Eq.~(\ref{Eq_avgN}) to $N+1/2$ and solving for $V$, we obtain an equation for the electron charging transition as a function of magnetic field. Setting $E_c$ and $\mu$ to zero (as they only cause an overall shift in $V$, which we define as $V_0$), the remaining fitting parameters are now $\alpha$, the electron temperature, and the various excited state energies of the polarized spin states. For the 1 electron to 2 electron charging transition, there are 2 and 4 possible spins states, respectively. Defining the singlet-triplet splitting of the 2 electron system as $E_{ST}$, we can now express the voltage where the 1-2 electron charging transition occurs as
\begin{multline}
V_{1 \rightarrow 2}(B) = V_0 + \\ \frac{1}{\alpha \beta}\log \left(\frac{\left(e^{\beta g \mu_B B}+1\right) e^{ \beta (
\frac{1}{2}g \mu_B B +E_{ST})}}{e^{\beta (g \mu_B B +E_{ST})}+e^{2 \beta g \mu_B B}+e^{ \beta g \mu_B B}+1}\right).
\label{eq_12MS}
\end{multline}

To account for the magnetic field dependence of the transverse component of the electron wave function in the dot and the chemical potential of the reservoir \cite{Stern:1968p1687,Weis:1993p4019,Weis:1994p664}, a quadratic term in magnetic field, $\gamma B^2$, is added to the 3-4 electron charging transition fitting function to better fit the magnetospectroscopy curves at large magnetic field.

The gate voltage where the 0-1, 2-3, and 3-4 electron charging transitions occur are
\begin{equation}
V_{0 \rightarrow 1}(B) = -\frac{\beta B g \mu_B +2 \log \left(e^{-\beta g \mu_B B}+1\right)}{2 \alpha \beta } + V_0,
\end{equation}
\begin{widetext}
\begin{equation}
V_{2 \rightarrow 3}(B) = \frac{1}{\alpha \beta}\log \left(\frac{e^{\beta (\frac{1}{2} g \mu_B B - E_{ST}+ 
 E_{S=3/2})} \left(e^{\beta (g \mu_B B + E_{ST})}+e^{\beta g \mu_B B}+e^{2\beta g \mu_B B}+1\right)}{(e^{\beta g \mu_B B}+1)(2e^{\beta(g \mu_B B + E_{S=3/2})} + e^{2 \beta g \mu_B B}+1)}\right)+V_0,
\end{equation}
\begin{multline}
V_{3 \rightarrow 4}(B) = \\ \frac{1}{\alpha \beta} \log \left(\frac{\left(e^{\beta g \mu_B B}+1\right) \left(2
 e^{\beta (g \mu_B B +E_{S=3/2})}+e^{2 \beta g \mu_B B}+1\right)
 e^{\beta (\frac{1}{2} g \mu_B B -E_{S=3/2}+E_{S=1}+
 E_{S=2})}}{d}\right) + \gamma B^2 + V_0,
\end{multline}
where the denominator is given by
\begin{multline}
d = 2 e^{\beta (2 g \mu_B B + E_{S=1}+ E_{S=2})}+e^{\beta (g \mu_B B + E_{S=1})} +e^{\beta(2 g \mu_B B + E_{S=1})}+e^{\beta (3 g \mu_B B + E_{S=1})}+e^{\beta(4 g \mu_B B + 
 E_{S=1})}+e^{\beta E_{S=1}} \\ +3 e^{\beta (g \mu_B B + E_{S=2})}+3 e^{\beta(2 g \mu_B B + 
 E_{S=2})}+3 e^{\beta(3 g \mu_B B + E_{S=2})},
\end{multline}
\end{widetext}
$E_{S=3/2}$ is the 3-electron excitation energy, and $E_{S=1}$ and $E_{S=2}$ are the 4-electron excitation energies.

As explained in the main text, the lower bounds of the valley and orbital splittings can be estimated from the zero-magnetic-field excited spin energies. Measuring at the 3-4 electron charging transition, the lowest lying valley splitting $E_{V1}$, orbital splitting $E_{Orb}$, and valley splitting $E_{V2}$ of the first excited orbital are given by
\begin{equation}
E_{V1} \geq E_{S=3/2}-E_{S=1},
\label{eq:v1}
\end{equation}
\begin{equation}
E_{Orb} \geq \frac{1}{2} E_{S=2},
\label{eq:orb}
\end{equation}
\begin{equation}
E_{V2} \geq E_{S=2}-E_{S=3/2}-E_{S=1}.
\label{eq:v2}
\end{equation}

\section{\textit{\lowercase{g}}-factor Measurement}
\label{A_gfactor}

\begin{figure}
\includegraphics[width=0.45\textwidth]{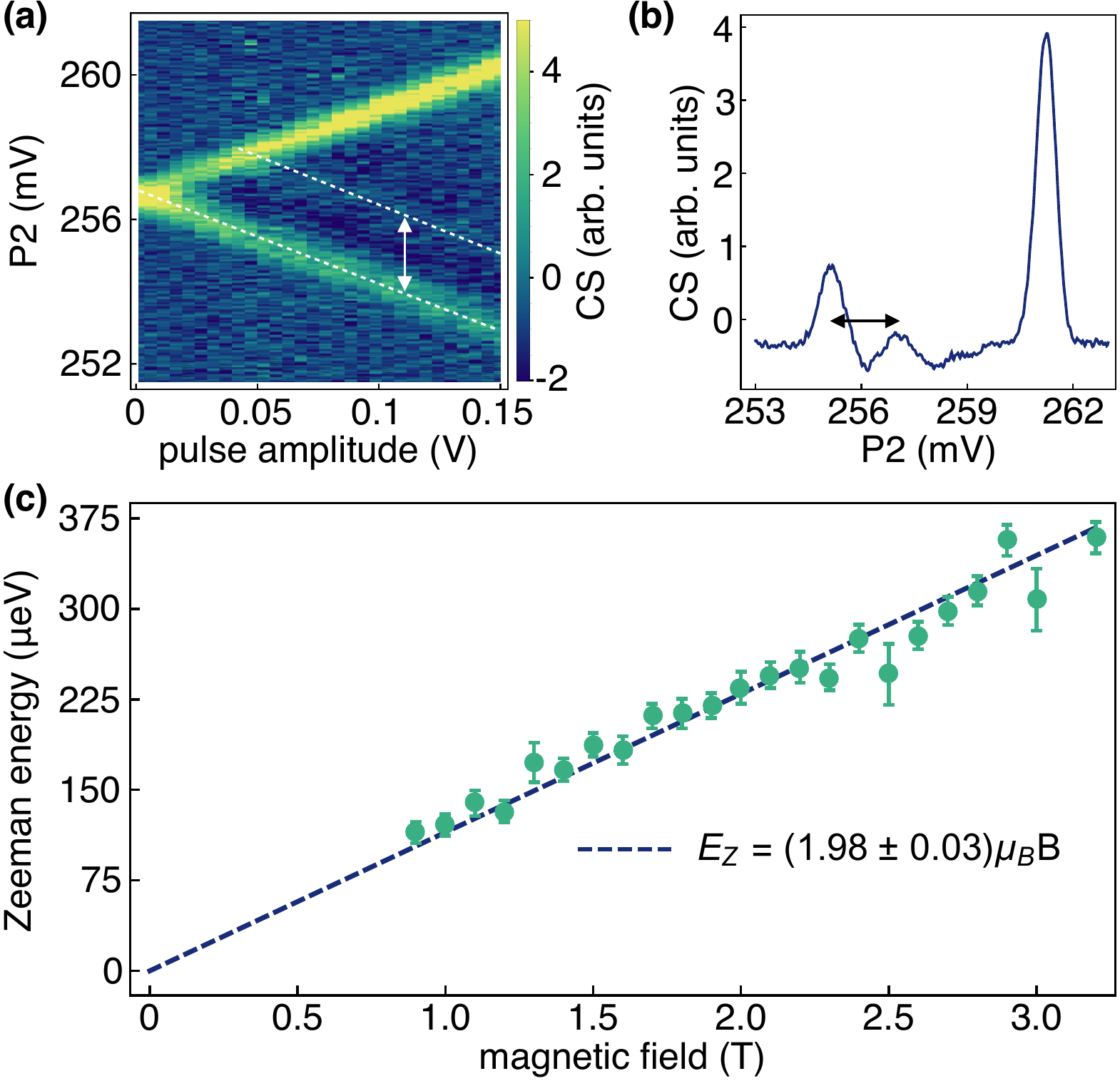}
\caption{
Electron $g$-factor measurement. (a) Pulsed-gate spectroscopy of the P2 quantum dot at the 0-1 electron charging transition. A $100-300$~\si{\kilo \hertz} square wave voltage tone is applied to the dot gate. A 2 Tesla in-plane magnetic field is applied to split the spin-dependent loading lines (highlighted with white dashed lines). The white arrow indicates the spin-splitting. (b) A highly averaged voltage scan at a fixed pulse amplitude and a 1.8 Tesla magnetic field. The black arrow indicates the spin-split loading lines. (c) The spin-splitting (Zeeman energy) is calculated from the voltage splitting, multiplied by the dot-to-gate lever arm, and is plotted as a function of applied magnetic field (green points). A linear fit (dashed line) through the data points yields the extracted $g$-factor.
}
\label{fig_gfactor}
\end{figure}

Measurements are performed to determine whether the electron $g$-factor is modified by the Ge spike or changes with magnetic field, as has been previously reported~\cite{Simmons:2011p156804}. Measuring the $g$-factor is also important for confirming the accuracy of energy splittings measured with magnetospectroscopy. Here, we employ a method similar to Ref.~\onlinecite{Simmons:2011p156804}, where pulsed-gate spectroscopy~\cite{Elzerman:2004p731,Hanson:2007p1217,Prance:2012p046808} was performed at various applied magnetic fields. Figure~\ref{fig_gfactor}(a) shows how the spin loading lines are split by the applied magnetic field. This spin splitting is tracked as a function of applied magnetic field. Using a lever-arm calculated through measurements of the thermal broadening of the detected charge sensor signal, this spin splitting in voltage space is converted to an energy splitting. Fitting these results linearly, as shown in Fig.~\ref{fig_gfactor}(c), we see that the calculated $g$-factor falls within the uncertainty of the expected value of $g \approx 2$~\cite{Wilamowski:2003}. 

\section{Electric Field Tuning of the states in the Quantum Dot}
\label{A_E_tuning}

\begin{figure}
\includegraphics[width = 0.45\textwidth]{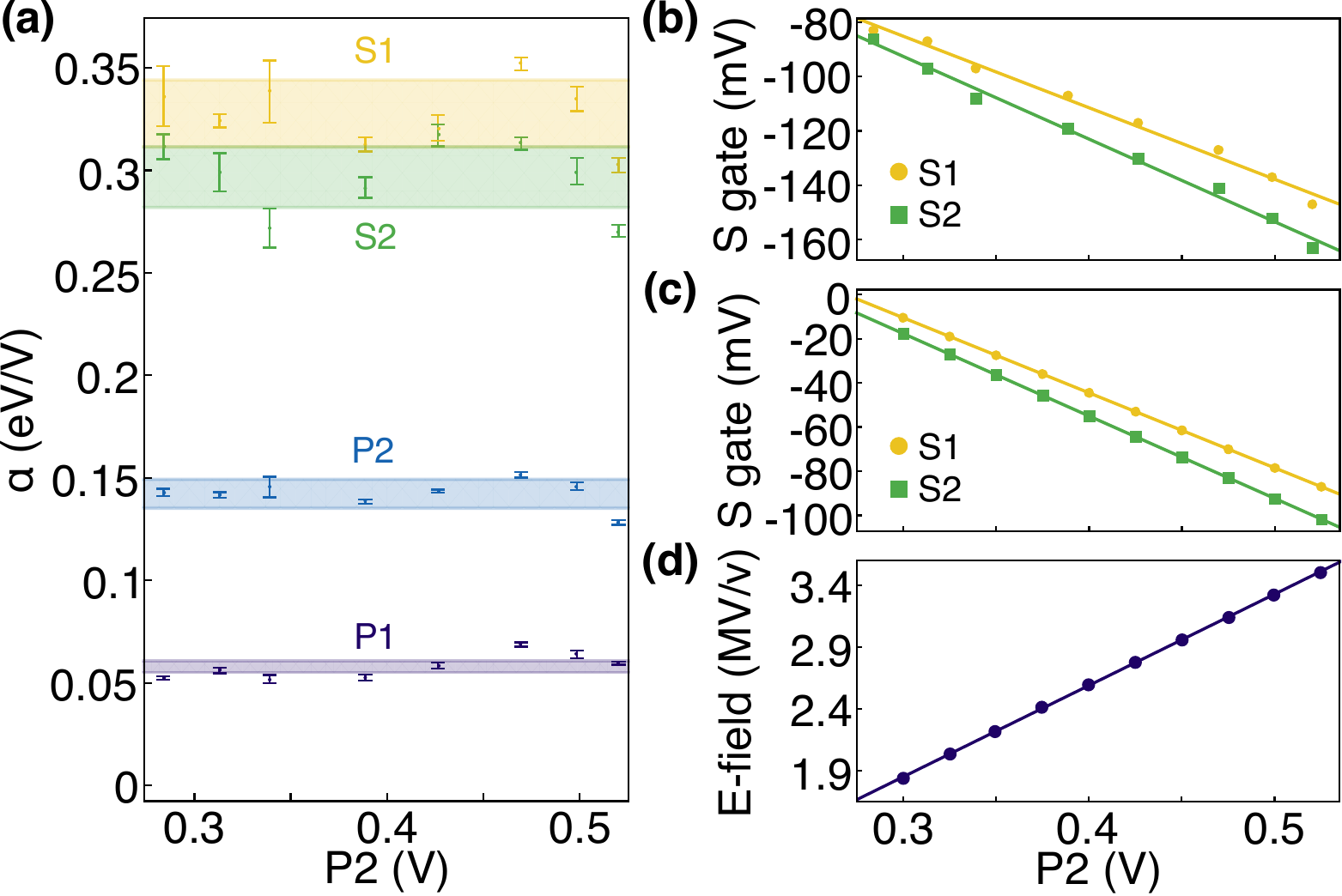}
\caption{
Gate tuning for experimental and simulated electric field tunings. (a) Lever arm ($\alpha$) calculations of the four nearby gates with the largest action on the quantum dot as a function of the experimental voltage tuning for Fig.\ \ref{fig_magspec}(e). See Fig.\ \ref{fig_device}(c) for gate labeling. The shaded area shows $\pm$5$\%$ variations around the weighted mean. (b) Experimental voltage tuning of the S1 and S2 gates, based on the relative lever arm difference found in (a). Lines are linear fits through the data. (c) Voltage tuning of S1 and S2 gates in a COMSOL simulation meant to extract the applied electric field at the location of the quantum dot. Relative tuning of S gates is based on the relative mutual capacitance between the S gate and the quantum dot. Lines are linear fits through the data. (d) Electric field results from the COMSOL simulated tuning in (c), showing a linearly increasing electric field as a function of P2 voltage. These electric field results are used for the tight-binding calculations shown in Fig.~\ref{fig_sim}(b).
}
\label{figStuning}
\end{figure}

\begin{figure}[t]
\includegraphics[width=0.45\textwidth]{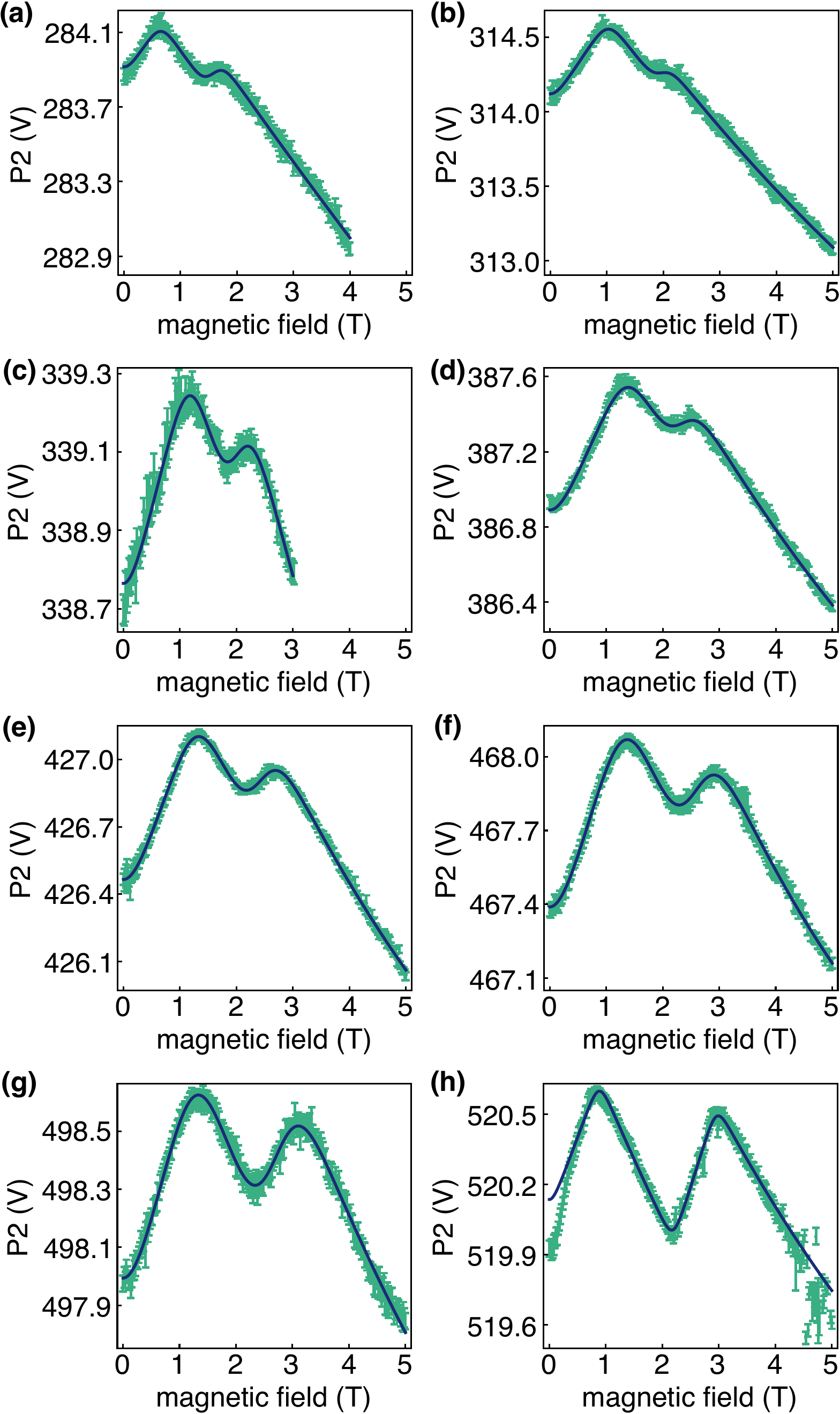}
\caption{
Electric field tuned magnetospectroscopy at the 3-4 electron charging transition. (a)-(h) Magnetospectroscopy results (green) across different gate tunings after registration of individual scans. Dark blue lines are fits to these data. Zero-magnetic-field spin splittings calculated from these scans are shown in Fig.~\ref{fig_magspec}(e).
}
\label{figSfits}
\end{figure}

The gates S1 and S2, labeled in Fig.~\ref{fig_device}(c), are used to modify the vertical electric field at the quantum dot. Relative lever arms of the S gates with respect to the plunger gate P2 are calculated from the slope of the electron charging transition in 2D plots of P2 and each S gate. The S gates are then tuned according to these lever arms such that the change in voltage of each of these gates affects the dot transition equally. Lowering the voltage on the S gates necessarily results in an increased P2 voltage to maintain the same charge occupation. Additionally, the barrier gates surrounding P2 are modified to maintain an adequate tunnel rate. Figure~\ref{figStuning}(a) shows the lever arm at each tuning for the 4 gates with the largest lever arms. The P2 lever arm is calculated from the magnetospectroscopy fit and the three other lever arms are calculated from the relative lever arms measured during tuning. The shaded area is $\pm 5 \%$ around the weighted mean, showing that the lever arms, and therefore dot positioning with respect to these gates, changes very little across the tuning range. The voltages of the S gates at each tuning are shown in Fig.~\ref{figStuning}(b).

This tuning scheme is replicated with Thomas-Fermi calculations using COMSOL multiphysics to estimate the electric field at the location of the dot and validate our interpretation of these voltage changes. The three aluminum gate layers are modeled using the intended fabrication dimensions and a 2DEG is located $2$~\si{\nano \meter} beneath the upper interface of a Si/SiGe interface. With P2 set to $300$~\si{\milli \volt} (lower end of experimental tunings), the nearest barrier and screening gate voltages are adjusted to achieve 3.5 electrons confined within the quantum dot. All other gate voltages are set to accumulate a charge density of $4 \times 10^{11}$~\si{\centi \meter}$^{-2}$. Similar to the experimental tuning scheme, the mutual capacitances between the dot and gates S1 and S2 are calculated. With the constraints of maintaining 3.5 electrons in the dot and the experimental P2 voltage range, S1 and S2 are `symmetrically' tuned according to their capacitance values. This has the effect of preserving the simulated lateral dot position to within $1$~\si{\nano \meter}. The resulting tuned screening voltages are shown in Fig.~\ref{figStuning}(c). Lack of simulated trapped charge and possible fabrication imperfections causes different voltage ranges to be obtained in the experimental and simulated screening voltage values. Figure~\ref{figStuning}(d) shows the calculated COMSOL electric field results at the center of the quantum dot, which are found to be linear and increasing with P2 gate voltage, as expected. These electric field results are used to define the electric field range in Fig.~\ref{fig_sim}(b).

Figure~\ref{figSfits} shows the extracted magnetospectroscopy curve for each tuning in Fig.~\ref{figStuning}. The zero-magnetic-field excited spin states calculated from the fits are shown in Fig.~\ref{fig_magspec}(e).

\section{Envelope functions for the tight-binding simulations}
\label{A_TB_functions}
In this Appendix, we compute the envelope functions for the two lowest valley states shown in Fig.~\ref{fig_sim}(a).

As noted in the main text and in Ref.~\cite{Friesen:2007p115318}, the wavefunction envelopes are modulated by fast oscillations, embodied in the factors $\cos(k_0z+\varphi)$ or $\sin(k_0z+\varphi)$, where $\pm k_0\hat z$ are the positions of the conduction valley minima in reciprocal space and $\varphi$ is the valley phase.
In the simplest case, both valley states have the same envelope.
However, a single-atom Ge spike causes strong hybridization of the quantum well subband orbitals, resulting in very different envelopes for the two valley states.
There is strong motivation for such hybridization because the ground state can significantly minimize its energy by choosing a phase $\varphi$ such that $\cos(k_0z+\varphi)=0$ at the location of the Ge spike, thus minimizing its potential energy.
Consequently, the ground state does not ``feel" the barrier, and its envelope takes the shape of an electron in a quantum well with no spike.
The opposite is true for the excited valley state, since both valley states share the same $\varphi$; in this case, the fast oscillations of the excited states are maximized at the location of the spike: $\sin(k_0z+\varphi)=1$.

We can model the effect of valley-orbit coupling in this system as follows.
The Hamiltonian for an electron in a quantum well, not including the Ge spike, is given by $H_0=-(\hbar^2/2m^*)\partial_z^2-eEz+V_\text{QW}(z)$, where $m^*$ is the longitudinal effective mass for Si, $E$ is the vertical electric field, and $V_\text{QW}(z)$ is the quantum well confinement potential.
We also consider a single-atom Ge spike potential, given by $V_s(z)=v_s\delta(z-z_s)$, where $\delta(z-z_s)$ is a Dirac delta function describing the spike, centered at $z=z_s$.
We must then solve the coupled set of Schr\"odinger equations~\cite{Friesen:2007p115318}
\begin{equation}
\sum_{j=\pm 1}\alpha_je^{ij(k_0z+\varphi)}(H_0+V_s-\epsilon)F_j(z) =0, \label{eq:coupled}
\end{equation}
where $(\alpha_-,\alpha_+)$ is the energy eigenvector and $\epsilon$ is the energy eigenvalue.
Normally, we would expect $V_\text{QW}$ to be the source of the valley splitting; however, here, the effect of $V_s$ is much stronger than $V_\text{QW}$, because it is located at a position where $F_j(z)$ is a maximum.
For an approximate solution, we therefore adopt $V_s$ as the valley-splitting potential.

Normally, $\alpha_-$, $\alpha_+$, and $\varphi$ would be determined by explicitly diagonalizing Eq.~(\ref{eq:coupled}).
However, to a very good approximation, we already know that the solution is given by $\alpha_{\pm}=1/\sqrt{2}$ for the ground state and $\alpha_{\pm}=\pm 1/\sqrt{2}$ for the excited state, with $\varphi$ chosen such that $\cos(k_0z_s+\varphi)=0$.
Hence, we can write the (now) decoupled Schr\"odinger equations as
\begin{gather}
\sqrt{2}\cos(k_0z+\varphi)(H_0+V_s-\epsilon)F_+(z)=0 , \\
\sqrt{2}\sin(k_0z+\varphi)(H_0+V_s-\epsilon)F_-(z)=0 .
\end{gather}
Because of the great difference in characteristic length scales associated with the fast oscillations in $\cos(k_0z+\varphi)$ and the much slower variations of $F_+(z)$, we may treat them independently.
(This is one possible statement of the effective mass approximation~\cite{Bastard:1988wave}.)
We therefore left-multiply the Schr\"odinger equations by $\sqrt{2}\cos(k_0z+\varphi)$ or $\sqrt{2}\sin(k_0z+\varphi)$ and integrate over a unit cell, to remove the fast oscillations, obtaining
\begin{gather}
(H_0-\epsilon)F_+(z)=0 , \\
(H_0+2V_s-\epsilon)F_-(z)=0 .
\end{gather}
These results are consistent with our previous claim that the ground state ($F_+$) does not feel the spike, while the excited state ($F_-$) feels a doubly tall barrier.
Separately solving these two equations yields the envelopes shown in Fig.~\ref{fig_sim}	(a). 
The full tight-binding solutions, which are also shown in that figure, are very well described by these envelopes, demonstrating that the approximations used here are very good.

\end{document}